\documentclass[11pt,a4paper]{article}
\usepackage[utf8]{inputenc}
\usepackage[T1]{fontenc}
\usepackage[english]{babel}
\usepackage{geometry}
\usepackage{graphicx}
\usepackage{amsmath,amssymb}
\usepackage{booktabs}
\usepackage{array}
\usepackage{float}
\usepackage{xcolor}
\usepackage{titlesec}
\usepackage{cite}
\usepackage{hyperref}
\definecolor{arxivblue}{RGB}{0,0,180}
\hypersetup{colorlinks=true,linkcolor=arxivblue,citecolor=arxivblue,urlcolor=arxivblue}
\renewcommand{\thesection}{\Roman{section}}
\renewcommand{\thesubsection}{\Alph{subsection}}
\titleformat{\section}{\normalfont\large\bfseries\centering}{\thesection.}{0.75em}{\MakeUppercase}
\titleformat{\subsection}{\normalfont\normalsize\bfseries}{\thesubsection.}{0.65em}{}
\titlespacing*{\section}{0pt}{2.7ex plus 1ex minus .2ex}{1.4ex plus .2ex}
\titlespacing*{\subsection}{0pt}{2.2ex plus .8ex minus .2ex}{0.9ex plus .2ex}
\title{\textbf{Optical-Fiber Activation of Observer-Split Source Terms in a Rotating Aluminum Disc}}
\author{A. Iadicicco$^{1}$\textsuperscript{\hyperlink{corrnote}{\textcolor{arxivblue}{a}}} and L. Verolino$^{2}$\\[0.55em]
$^{1}$\textit{Universita Popolare ``Nikola Tesla'',}\\
\textit{Via Trinita 33, 83100 Avellino, Italy}\\[0.25em]
$^{2}$\textit{Universita di Napoli ``Federico II'',}\\
\textit{Dipartimento di Ingegneria Elettrica e delle Tecnologie dell'Informazione,}\\
\textit{Via Claudio 21, 80125 Napoli, Italy}}
\date{}
\renewenvironment{abstract}{%
  \vspace{-1.35em}
  \noindent\hspace*{0.055\textwidth}%
  \begin{minipage}{0.89\textwidth}\small
}{%
  \end{minipage}%
  \par\vspace{0.55em}%
}
\begin{document}
\maketitle
\begingroup
\renewcommand{\thefootnote}{\alph{footnote}}
\footnotetext[1]{\hypertarget{corrnote}{}Corresponding author. Email address: {\hypersetup{urlcolor=black}\href{mailto:antonio.iadicicco@unitesla.eu}{\nolinkurl{antonio.iadicicco@unitesla.eu}}}}
\endgroup

\begin{abstract}
We propose an optical-fiber extension of the Observer-Split source framework introduced by Iadicicco, Modanese, and Verolino for finite rotating systems in extended Aharonov-Bohm electrodynamics. In the parent formulation, a covariantly conserved current is rewritten in a rotating 3+1 split, producing the exact observer-adapted source term \(I_{\rm split}=N^{-1}D_i(\rho\beta^i)\). In the weak-field rigid-rotation limit this becomes \(I_G=\Omega\partial_\phi\rho\), so a localized angular density transported by the apparatus generates a sign-reversing effective source. The present work replaces the magnetic source proxy used in historical rotating-magnetic-system benchmarks, including the Roshchin-Godin report, with an optically gated aluminum-induced source density carried by a rotating aluminum disc. The evanescent/tunneling optical coefficient is not identified with a charge density. Instead, it is treated as a controllable gate that modulates an effective surface or active-layer density induced in the aluminum carrier. Conditional on the calibration of this density, the proposed source coordinate is
\[
\Delta G_{\rm opt}=\Omega\,\chi_{\rm Al}\,T_{\rm ev}e^{-1/2}/\sigma .
\]
The residual vertical response is modeled through the experimentally identifiable product \(\Lambda_{\rm opt}=\kappa_{\rm opt}\chi_{\rm Al}\), not by assigning an uncalibrated force in newtons before measurement. The paper therefore gives the sensitivity target, the minimum detectable force \(F_{\rm min}\), the null bound on \(|\Lambda_{\rm opt}|\), and the inverse calibration law by which \(\chi_{\rm Al}\) and \(\kappa_{\rm opt}\) can be separated after independent density calibration. The proposed demonstrator uses an illuminated fiber-optic ring masked to a 10--20 degree active sector, ON/OFF optical gating, CW/CCW reversal, material controls, dummy controls, quantitative null bounds, and a progressive validation ladder. An extended attribution path is then specified for deciding whether a positive device-level signature can be promoted to evidence of a gravitational interaction: external test masses, free-fall readout, distance and orientation laws, material-independence tests, shielding, vacuum, and independent replication. A synthetic neural-network audit is used only as an internal consistency test showing that the modeled signature is mathematically separable from declared synthetic false-positive classes. The work is a falsifiable theoretical and experimental protocol for testing anomalous gravitational-response signatures; it reports no direct anomalous-weight measurements and makes no claim of evidence for a gravitational anomaly or for extended Aharonov-Bohm electrodynamics.
\end{abstract}

\section{Introduction}
Rotating finite devices provide a natural laboratory in which observer-adapted transport variables can differ from covariant microscopic conservation variables. This work is conceived as a direct optical continuation of the Observer-Split model developed by Iadicicco, Modanese, and Verolino \cite{iadicicco2026split}, in which observer-adapted source terms arise from the rotating 3+1 decomposition of a covariantly conserved current. The parent framework starts from the no-go statement that standard generally covariant, locally \(U(1)\)-invariant matter does not acquire a true microscopic charge anomaly from rotation alone. Nevertheless, when the conserved current is described in a rotating 3+1 slicing, the transport continuity equation contains an exact observer-adapted source term.

The present paper asks whether the active source in such a rotating finite device must be magnetic, or whether an optical-fiber evanescent/tunneling gate can modulate an aluminum-induced effective density while preserving the same observer-split source coordinate. This is motivated by the mathematical analogy between electronic tunneling and optical evanescent tunneling: both contain exponentially decaying barrier modes and a transmission coefficient controlling an effective boundary-mediated transfer \cite{simmons1963,hartman1962,steinberg1993,balcou1997,longhi2005,philbin2008}. The optical coefficient is therefore not used as a charge density by itself; it is used as a controllable gate acting on the metallic carrier.

The manuscript is intentionally formulated as a falsifiable protocol. A first positive ON/OFF and CW/CCW weight signature would not by itself establish extended Aharonov-Bohm electrodynamics or a gravitational origin for the observed anomaly. It would instead be a candidate anomalous force signature. The proposed validation ladder then tests whether the signal survives material, thermal, vibrational, aerodynamic, optical-power, orientation, and independent-replication controls.

\section{Historical rotating magnetic-system benchmark}
Roshchin and Godin reported anomalous weight variations in a rotating magnetic system \cite{roshchin2000,roshchin2004}; the present work proposes an optical Observer-Split route for studying possible anomalous gravitational-response signatures using optical fiber and a rotating aluminum disc. The reported phenomenology included an onset region, a maximum apparent weight reduction near a critical rotational regime, and reversal of the observed effect when the direction of rotation was inverted. These results remain historically important and require modern independent replication under controlled metrology. Here they are used only as a benchmark for finite-device rotating-source scaling, not as proof.

In that benchmark class, one may encode the active finite-device coordinate as
\begin{equation}
\Delta G_{\rm mag}=\Omega B_{\rm eff}e^{-1/2}/\sigma_{\rm mag},
\end{equation}
where \(B_{\rm eff}\) is an operational magnetic source amplitude and \(\sigma_{\rm mag}\) is the angular source width. The optical construction below preserves the sign-reversal and finite-source scaling logic, but it does not equate a magnetic field directly with an optical transmission coefficient. Instead, the optical gate is allowed to modulate a phenomenological aluminum-induced density.

\section{Observer-split source structure}
Consider the ADM metric, in the rotating-frame and gravito-electromagnetic context discussed in Refs.~\cite{iadicicco2026split,ruggiero2005},
\begin{equation}
ds^2=-N^2dt^2+h_{ij}(dx^i+\beta^idt)(dx^j+\beta^jdt),
\end{equation}
and decompose the current as \(J^\mu=\rho n^\mu+j^\mu\). Covariant conservation \(\nabla_\mu J^\mu=0\) gives, in observer-adapted transport variables,
\begin{equation}
\partial_t\rho+D_i(Nj^i)=D_i(\rho\beta^i).
\end{equation}
The exact split source is therefore
\begin{equation}
I_{\rm split}=N^{-1}D_i(\rho\beta^i).
\end{equation}
For weak-field rigid rotation, \(\beta=\Omega\times r\), and
\begin{equation}
I_G=(\Omega\times r)\cdot\nabla\rho=\Omega\,\partial_\phi\rho .
\end{equation}
A uniform or perfectly axisymmetric source is suppressed. A localized optical sector satisfies
\begin{equation}
\rho_{\rm eff}(\phi,t)=\rho_0+\delta\rho_s(\phi-\Omega t),
\qquad
I_G\simeq\Omega\,\partial_\phi\delta\rho_s .
\end{equation}

\begin{figure}[H]
\centering
\includegraphics[width=0.88\linewidth]{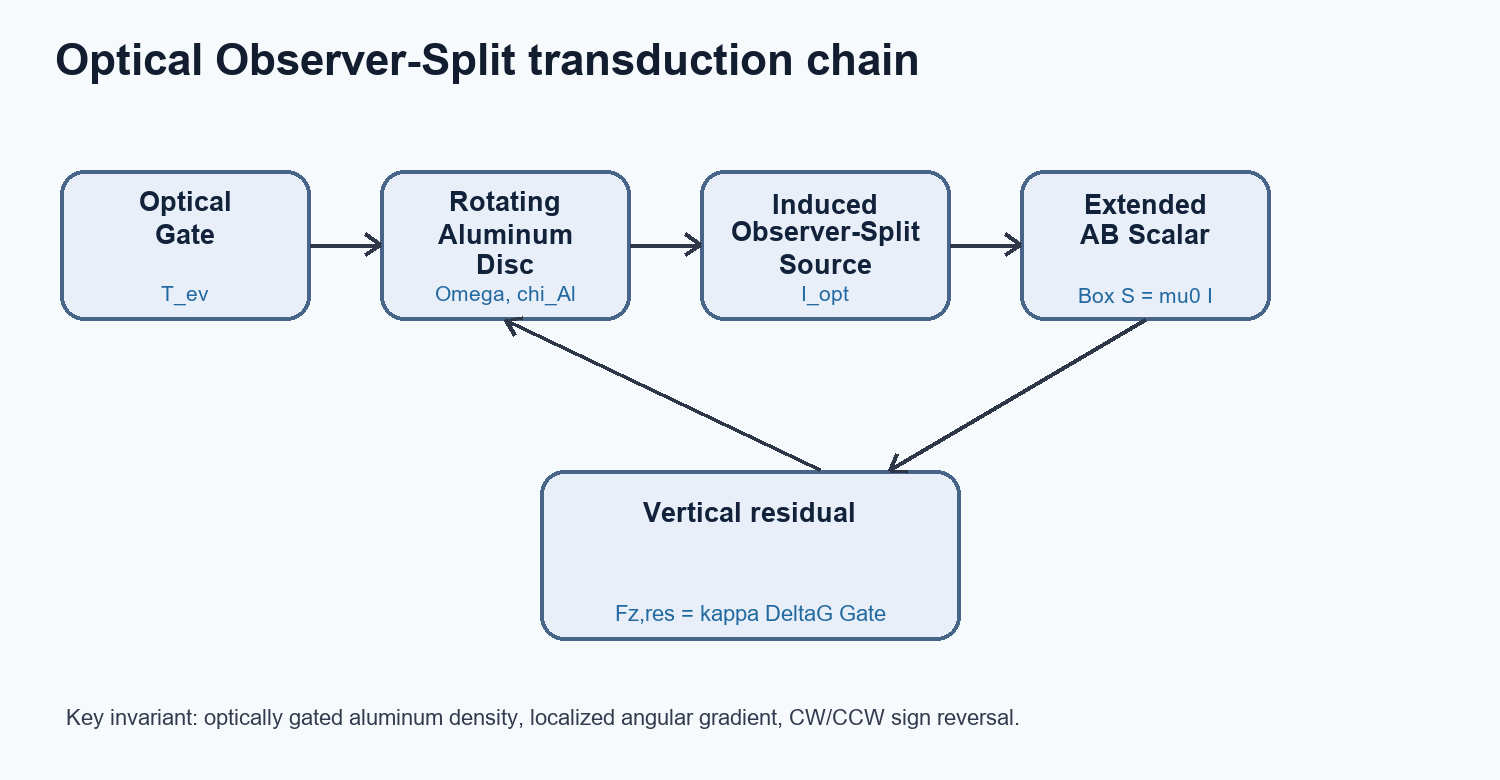}
\caption{The optical-fiber transduction chain. Evanescent optical gating modulates an aluminum-induced effective density while preserving the observer-split source coordinate.}
\end{figure}

\section{Reference frames and transported source profile}
The induced density used in the experiment is defined operationally in the local body frame of the aluminum disc, where the illuminated sector is stationary. Let \((t,r,\phi,z)\) be laboratory cylindrical coordinates and let
\begin{equation}
\theta=\phi-\Omega t-\phi_0
\end{equation}
be the angular coordinate comoving with the active optical sector. The source profile may then be written as
\begin{equation}
\rho_{\rm ind,Al}(t,r,\phi,z)
=\rho_b+\delta\rho_{\rm Al}(r,z)\,f(\theta),
\end{equation}
where \(f\) is the measured angular optical profile and \(\delta\rho_{\rm Al}\) is the calibrated aluminum response amplitude. In the laboratory description,
\begin{equation}
\partial_t\rho_{\rm ind,Al}=-\Omega\,\partial_\phi\rho_{\rm ind,Al}.
\end{equation}
The Eulerian observers associated with the ADM split measure the transport density entering \(I_{\rm split}=N^{-1}D_i(\rho\beta^i)\). For a small demonstrator with \(v=\Omega r\ll c\), the difference between the local comoving density and the corresponding laboratory density is of order \(v^2/c^2\), while the leading observer-split term is the angular transport term
\begin{equation}
I_{\rm opt}\simeq\Omega\,\partial_\phi\rho_{\rm ind,Al}.
\end{equation}
This makes the frame assignment explicit: the optical sector is fixed in the rotating material body, the phase gradient is evaluated as a transported profile in laboratory angular coordinates, and the sign change arises from reversing the transported angular velocity \(\Omega\).

\section{Optical evanescent tunneling bridge}
For guided light, a scalar component of the field satisfies a Helmholtz equation of the form
\begin{equation}
\left[\nabla_\perp^2+k_0^2n^2(x)-\beta_{\rm opt}^2\right]E(x)=0 .
\end{equation}
In a classically forbidden or lower-index region the field is evanescent:
\begin{equation}
E_{\rm ev}(x)=E_0e^{-\alpha_{\rm ev}x},
\qquad
\alpha_{\rm ev}=k_0\sqrt{n_{\rm core}^2\sin^2\theta-n_{\rm out}^2}.
\end{equation}
The optical tunneling or evanescent coupling coefficient may be written operationally as
\begin{equation}
T_{\rm ev}(g,\lambda)=T_0e^{-2\alpha_{\rm ev}g}.
\end{equation}
This is not an electronic current, and it is not a charge density. Rather, it is a photon/field tunneling coefficient used here as a controllable optical gate. The physical source density entering the Observer-Split term is taken to be an effective density induced or modulated in the aluminum carrier, with the optical field acting as the switchable angular control.

\section{Optically gated aluminum-induced source coordinate}
Let the fiber ring be masked so that only a localized angular sector is optically active:
\begin{equation}
T_{\rm ev}(\phi,t)=T_0\exp\left[-\frac{(\phi-\Omega t-\phi_0)^2}{2\sigma^2}\right].
\end{equation}
The optical coefficient is not identified with \(\rho\). Instead, define an aluminum-induced effective density
\begin{equation}
\rho_{\rm ind,Al}(\phi,t)=\rho_b+\chi_{\rm Al}(P_{\rm opt},\lambda,\sigma_{\rm Al},g,\mathcal{G})\,T_{\rm ev}(\phi,t),
\end{equation}
where \(\chi_{\rm Al}\) is a phenomenological transduction coefficient with the dimensions required for a density, \(P_{\rm opt}\) is optical power, \(\lambda\) is wavelength, \(\sigma_{\rm Al}\) denotes relevant aluminum material parameters, \(g\) denotes an evanescent-coupling scale when present, and \(\mathcal{G}\) represents disc and boundary geometry. This coefficient may include surface polarization, photo-induced charge redistribution, photothermal electronic response, ponderomotive surface effects, or any other calibrated optical-to-metal transduction channel. It is not fixed a priori by the evanescent coefficient.

The Observer-Split source is therefore built from the induced density:
\begin{equation}
I_{\rm opt}=\Omega\,\partial_\phi\rho_{\rm ind,Al}
=\Omega\,\chi_{\rm Al}\,\partial_\phi T_{\rm ev}.
\end{equation}
At one Gaussian width,
\begin{equation}
\max|\partial_\phi T_{\rm ev}|=T_0e^{-1/2}/\sigma ,
\end{equation}
so the measurable source coordinate is
\begin{equation}
\boxed{\Delta G_{\rm opt}=\Omega\,\chi_{\rm Al}T_0e^{-1/2}/\sigma .}
\end{equation}

For a generic carrier material, one would replace \(\chi_{\rm Al}\) by \(\chi_{\rm mat}\). The aluminum disc is therefore not a passive holder only; it is the rotating conductive transducer. An insulating plastic carrier corresponds to the control hypothesis \(\chi_{\rm plastic}\ll\chi_{\rm Al}\), and is expected to suppress the modeled Observer-Split response if the metallic transduction channel is essential.

\subsection{Effective density, current, and calibration status}
The induced aluminum density may be represented either as a surface density on the illuminated active layer,
\begin{equation}
\sigma_{\rm ind,Al}(\phi,t)=\sigma_b+\chi_{\rm Al}^{(s)}(P_{\rm opt},\lambda,\sigma_{\rm Al},g,\mathcal{G})T_{\rm ev}(\phi,t),
\qquad [\chi_{\rm Al}^{(s)}]={\rm C\,m^{-2}},
\end{equation}
or as an equivalent volume density over an effective active depth \(d_{\rm eff}\),
\begin{equation}
\rho_{\rm ind,Al}=\sigma_{\rm ind,Al}/d_{\rm eff},
\qquad
\chi_{\rm Al}=\chi_{\rm Al}^{(s)}/d_{\rm eff},
\qquad [\chi_{\rm Al}]={\rm C\,m^{-3}}.
\end{equation}
The sign of \(\chi_{\rm Al}\) is not assumed; it is part of the calibration and depends on the microscopic optical-to-metal channel. A complete microscopic derivation would require a physical current \(J^\mu_{\rm ind,Al}\). In the present phenomenological protocol this current is treated as an effective conserved carrier response,
\begin{equation}
J^\mu_{\rm ind,Al}=\rho_{\rm ind,Al}n^\mu+j^\mu_{\rm ind,Al},
\qquad
\nabla_\mu J^\mu_{\rm ind,Al}=0,
\end{equation}
so that the Observer-Split source used below is an observer-adapted transport term, not a microscopic nonconservation:
\begin{equation}
I_{\rm opt}=N^{-1}D_i(\rho_{\rm ind,Al}\beta^i).
\end{equation}
Operationally, \(\chi_{\rm Al}\) should be measured or bounded independently by a lock-in surface-potential, capacitive, or electrometer measurement synchronized to the optical ON/OFF modulation and rotor phase:
\begin{equation}
\chi_{\rm Al}^{\rm cal}
=\frac{\Delta\rho_{\rm ind,Al}^{\rm meas}}{\Delta T_{\rm ev}}.
\end{equation}
If no independent density is detected, the experiment still provides an upper bound on \(|\chi_{\rm Al}|\) for the tested optical geometry.

\begin{figure}[H]
\centering
\includegraphics[width=0.88\linewidth]{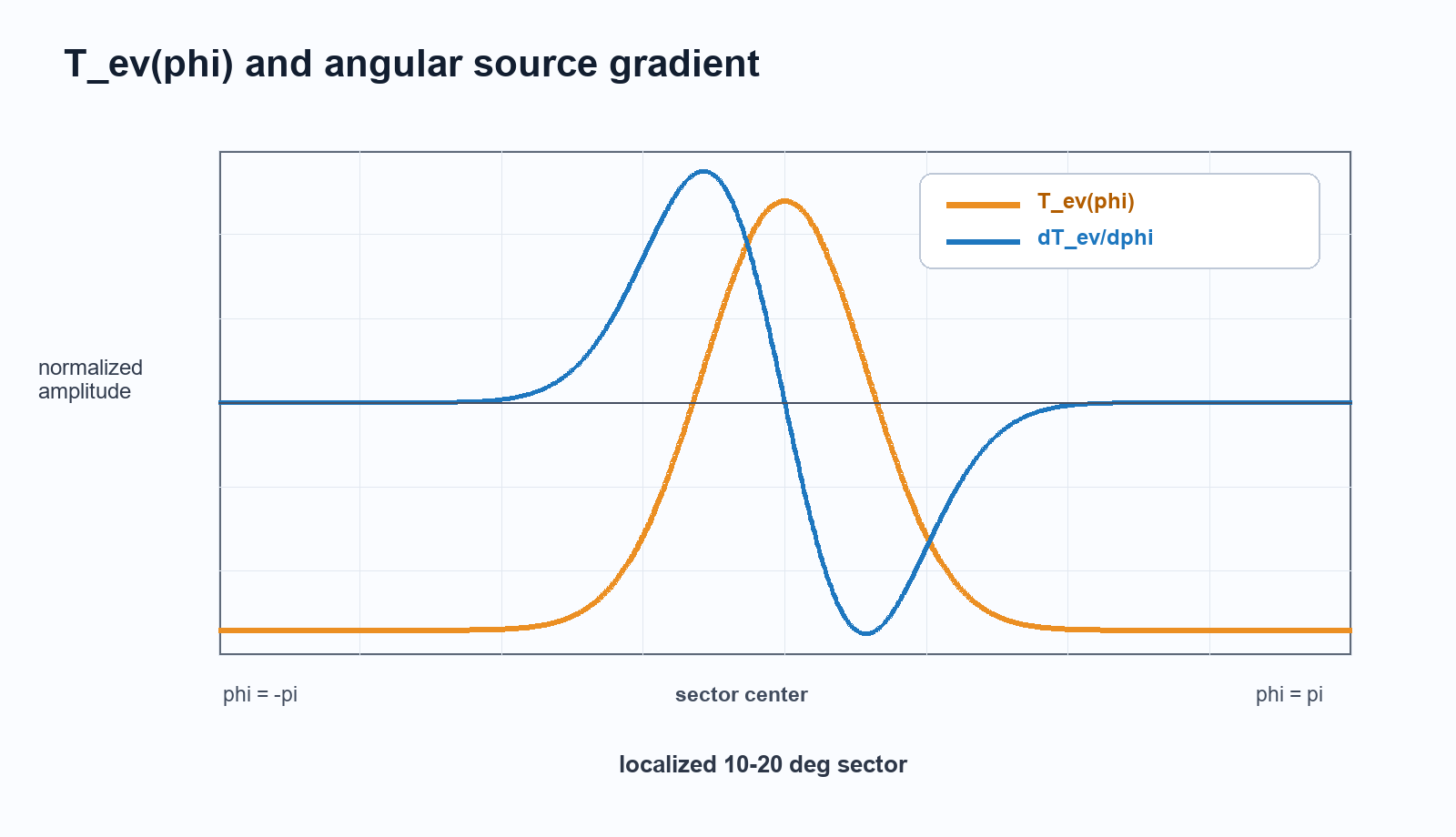}
\caption{A masked active optical sector creates a finite angular gradient. A fully uniform luminous ring is strongly suppressed in the \(\Omega\partial_\phi\rho_{\rm eff}\) channel.}
\end{figure}

\section{Conditional extended Aharonov-Bohm scalar closure}
This phenomenological construction is formulated to be compatible with the extended Aharonov-Bohm Observer-Split framework \cite{iadicicco2026split,modanese2017,minotti2021}, conditional on the existence and calibration of the proposed aluminum-induced effective density. The optical field is not treated as the source itself; it acts as a controllable gate that modulates \(\rho_{\rm ind,Al}\), from which \(I_{\rm opt}\) is constructed. The observer-split source can then be inserted into the extended Aharonov-Bohm scalar sector \cite{aharonov1959,aharonov1961,chambers1960,modanese2017,minotti2021} through
\begin{equation}
\nabla_\mu F^{\mu\nu}=J^\nu_{\rm eff}+i^\nu_{\rm split},
\qquad
i^\nu_{\rm split}=-\nabla^\nu\Box_g^{-1}I_{\rm opt},
\end{equation}
with
\begin{equation}
\Box_g S=\mu_0 I_{\rm opt}, \qquad S=\nabla_\mu A^\mu .
\end{equation}
The scalar equation alone does not uniquely determine a vertical force without a material response law. We therefore introduce a response functional
\begin{equation}
F^{\rm res}_z=\int_{\mathcal{V}}K_z(\mathbf{x};\Omega,\mathcal{G})\,S(\mathbf{x})\,d^3x ,
\end{equation}
where \(K_z\) is an apparatus-dependent coupling kernel to be calibrated or bounded. For the minimal demonstrator this is reduced to the phenomenological one-parameter source coordinate
\begin{equation}
F^{\rm res}_z=\kappa_{\rm opt}\Delta G_{\rm opt}
\left[1+\exp\left(-\frac{|\Delta G_{\rm opt}|-\Delta G_c}{w}\right)\right]^{-1}.
\end{equation}
Thus \(\kappa_{\rm opt}\), \(\Delta G_c\), and \(w\) are not presented as first-principles constants; they are experimental response parameters. Observation of the correct sign reversal constrains them, while a null result bounds the product \(|\kappa_{\rm opt}\chi_{\rm Al}|\) for the tested apparatus.
For pre-calibration planning it is useful to combine the two unknown transduction factors into
\begin{equation}
\Lambda_{\rm opt}\equiv\kappa_{\rm opt}\chi_{\rm Al}.
\end{equation}
In the low-activation or locally linear regime \(A\simeq1\), the odd force channel becomes
\begin{equation}
R_{\rm odd}\simeq
\Lambda_{\rm opt}\frac{\Omega T_0e^{-1/2}}{\sigma}.
\end{equation}
Consequently, a positive experiment identifies the reduced apparatus coefficient
\begin{equation}
\boxed{
\Lambda_{\rm opt}^{\rm exp}
=R_{\rm odd}\frac{\sigma}{\Omega T_0e^{-1/2}A}
}
\end{equation}
for each tested geometry. If the independent surface-density calibration gives \(\chi_{\rm Al}^{\rm cal}\neq0\), the vertical response coefficient is then separated as
\begin{equation}
\kappa_{\rm opt}^{\rm exp}
=\frac{\Lambda_{\rm opt}^{\rm exp}}{\chi_{\rm Al}^{\rm cal}}.
\end{equation}
This is the central identifiability rule of the protocol. Before calibration the experiment should not quote a force prediction in newtons as a theoretical certainty; after calibration it can either measure \(\Lambda_{\rm opt}\), separate \(\kappa_{\rm opt}\) and \(\chi_{\rm Al}\), or place a numerical upper bound on their product.
Under reversal of the rotating source,
\begin{equation}
\Omega\rightarrow-\Omega
\quad\Rightarrow\quad
\Delta G_{\rm opt}\rightarrow-\Delta G_{\rm opt}
\quad\Rightarrow\quad
F^{\rm res}_z\rightarrow-F^{\rm res}_z ,
\end{equation}
up to assembly sign conventions.

\begin{figure}[H]
\centering
\includegraphics[width=0.88\linewidth]{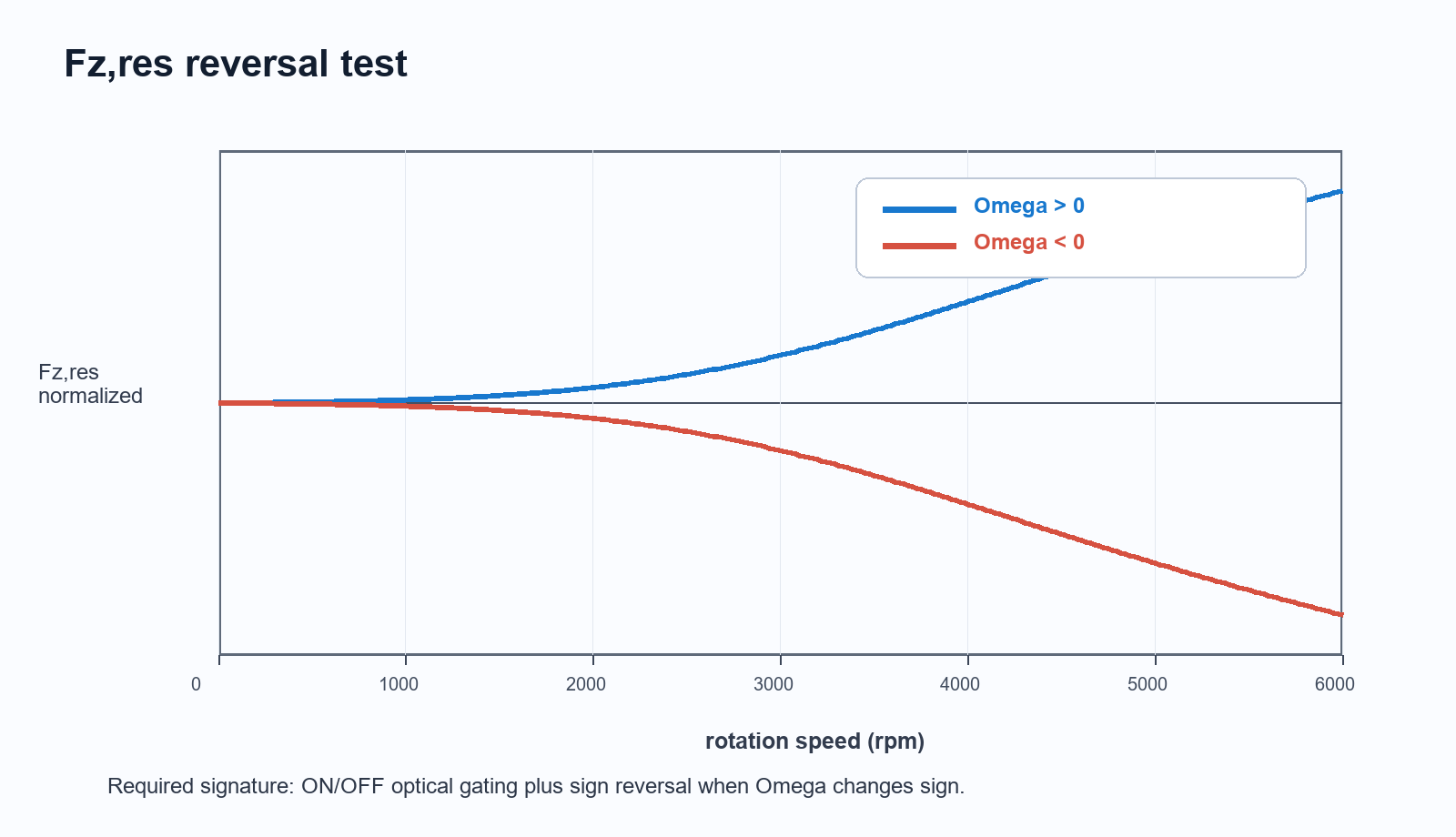}
\caption{Expected reversal signature. One rotation direction may show apparent weight reduction while the opposite direction shows apparent weight increase. The invariant prediction is sign reversal.}
\end{figure}

\section{Illustrative toy source-to-mass scaling}
The following comparison is a toy scaling index, not a first-principles prediction and not a direct equivalence between magnetic field amplitude and optical transmission. The optical gate is not expected to exceed a magnet in absolute field amplitude. Its possible advantage is source-to-mass efficiency, low rotor mass, clean ON/OFF gating, and reduced magnetic-attraction artifacts. Define
\begin{equation}
\mathcal{E}_{\rm opt}=\frac{\kappa_{\rm opt}\chi_{\rm Al}T_0}{\sigma_{\rm opt}M_{\rm opt}},
\qquad
\mathcal{E}_{\rm mag}=\frac{\kappa_{\rm mag}B_{\rm eff}}{\sigma_{\rm mag}M_{\rm mag}} .
\end{equation}
In this illustrative index, optical-fiber transduction would be favored when
\begin{equation}
\boxed{
\frac{\kappa_{\rm opt}\chi_{\rm Al}T_0}{\sigma_{\rm opt}M_{\rm opt}}
>
\frac{\kappa_{\rm mag}B_{\rm eff}}{\sigma_{\rm mag}M_{\rm mag}} .}
\end{equation}
Using the current small-rotor design comparison, the illustrative optical equivalent source amplitude needed to match the magnetic source-to-mass index is approximately
\begin{equation}
A_{\rm opt,match}\simeq 1.45\times10^{-2}.
\end{equation}
This value is a matching threshold inside the toy source-to-mass index. It is not a predicted weight-change fraction, not a calibrated optical power, and not an independently derived force scale. The actual residual amplitude depends on the experimentally realized product \(\kappa_{\rm opt}\chi_{\rm Al}\), the rotation rate, the angular localization, and the activation gate.

\begin{figure}[H]
\centering
\includegraphics[width=0.86\linewidth]{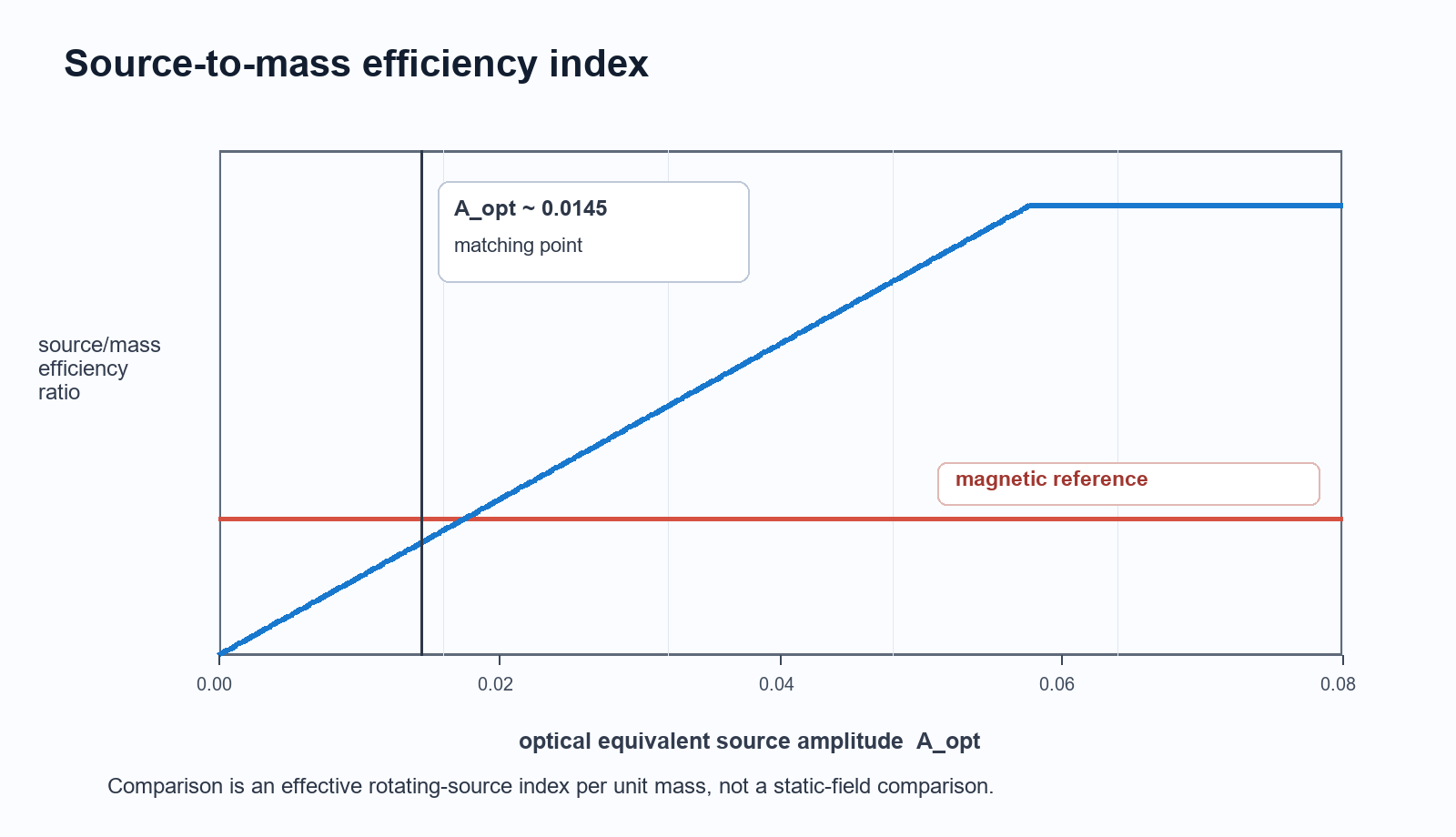}
\caption{Illustrative toy source-to-mass comparison. The plotted threshold is not a predicted weight fraction; it only compares effective rotating-source indices per unit mass.}
\end{figure}

\section{Minimal experimental demonstrator}
The experiment uses a 120 mm aluminum disc with a single fiber-optic side-glow ring. Most of the ring is covered with opaque tape; only a 10--20 degree sector remains luminous. The LED/controller and battery rotate with the disc. This prevents cable forces from contaminating weight measurements.

The primary acceptance condition is
\begin{equation}
\Delta W_{\rm CW}=W_{\rm CW,on}-W_{\rm CW,off},\qquad
\Delta W_{\rm CCW}=W_{\rm CCW,on}-W_{\rm CCW,off},
\end{equation}
with
\begin{equation}
\Delta W_{\rm CW}\simeq-\Delta W_{\rm CCW},
\qquad
\Delta W_{\rm dummy}\simeq0.
\end{equation}
Failure of ON/OFF gating, reversal, or dummy suppression falsifies the optical observer-split interpretation for that apparatus.

\section{Quantitative readout and null limits}
The primary measured quantity should be the component that is odd under rotation reversal and gated by light:
\begin{equation}
R_{\rm odd}=\frac{1}{2}\left(\Delta W_{\rm CW}-\Delta W_{\rm CCW}\right),
\qquad
C_{\rm even}=\frac{1}{2}\left(\Delta W_{\rm CW}+\Delta W_{\rm CCW}\right).
\end{equation}
The modeled signature requires \(R_{\rm odd}\neq0\) and \(C_{\rm even}\simeq0\) within the artifact budget. If a balance or force sensor has single-run standard deviation \(\sigma_W\) and \(N\) independent ON/OFF pairs are acquired, a conservative detection threshold may be defined as
\begin{equation}
F_{\rm min}=z_\alpha\frac{\sigma_W}{\sqrt{N}},
\end{equation}
where \(z_\alpha\) is the selected confidence factor. The experiment is therefore designed by declaring the sensitivity, not by inventing an uncalibrated force amplitude. A practical pre-registration table is:

\begin{table}[H]
\centering
\small
\begin{tabular}{>{\raggedright\arraybackslash}p{0.27\linewidth}>{\raggedright\arraybackslash}p{0.25\linewidth}>{\raggedright\arraybackslash}p{0.38\linewidth}}
\toprule
Quantity & Planned value or range & Role in the test \\
\midrule
Disc diameter & 120 mm & Minimal aluminum rotating carrier. \\
Active optical sector & 10--20 degrees & Creates the non-uniform angular gradient; uniform 360 degree light is the suppression control. \\
Effective angular width \(\sigma\) & \(0.07\)--\(0.15\) rad, fitted from the measured light profile & Enters \(\Delta G_{\rm opt}\propto1/\sigma\). \\
Rotation rate \(\Omega\) & 200, 500, 1000 rpm steps & Tests the predicted odd scaling with \(\Omega\). \\
Optical gate \(T_0\) & normalized measured ON/OFF amplitude, \(0<T_0\le1\) & Avoids assuming that LED brightness is equal to evanescent coupling. \\
Force noise \(\sigma_W\) & target \(10^{-5}\)--\(10^{-4}\) N per ON/OFF pair & Determines the measurable newton-scale threshold. \\
Number of pairs \(N\) & 100--400 per condition & Lowers \(F_{\rm min}\) as \(N^{-1/2}\). \\
Confidence factor \(z_\alpha\) & 3 for a first high-confidence screen & Defines the pre-registered detection threshold. \\
Example threshold & \(F_{\rm min}=3\times10^{-6}\) N for \(\sigma_W=10^{-5}\) N and \(N=100\) & Minimum odd force needed before claiming a candidate signature. \\
Conventional-force budget & each channel below \(F_{\rm min}/3\), total below \(F_{\rm min}\) & Required for accepting \(R_{\rm odd}\) as non-conventional. \\
\bottomrule
\end{tabular}
\caption{Planned operational parameters. Values are design targets to be replaced by measured calibration values in the laboratory log. They define the sensitivity of the experiment without pretending to know \(\chi_{\rm Al}\) or \(\kappa_{\rm opt}\) before calibration.}
\end{table}

For example, using \(F_{\rm min}=3\times10^{-6}\) N, \(\Omega=100\,{\rm s}^{-1}\), \(\sigma=0.10\), \(T_0=1\), and \(A=1\), the detectable reduced coefficient is
\begin{equation}
|\Lambda_{\rm opt}|_{\rm det}
\simeq
\frac{F_{\rm min}\sigma}{\Omega T_0e^{-1/2}A}
\simeq
5\times10^{-9}\ {\rm N\,s}
\end{equation}
in the reduced one-coordinate model. This number is not a predicted force; it is the planned sensitivity to the unknown reduced coefficient. A null result at rotation rate \(\Omega\) then gives the apparatus-level bound
\begin{equation}
|\kappa_{\rm opt}\chi_{\rm Al}|
<
\frac{F_{\rm min}\,\sigma}
{|\Omega|\,T_0e^{-1/2}\,A(|\Delta G_{\rm opt}|)} ,
\end{equation}
for the tested geometry and optical coupling. This is the quantitative meaning of a negative test: it constrains the product of phenomenological transduction and vertical response coefficients rather than falsifying the general Observer-Split identity.

The artifact budget should be reported next to \(R_{\rm odd}\). At minimum it includes eccentric mass forces \(F_{\rm imb}=m_{\rm ecc}r\Omega^2\), ordinary optical pressure \(F_{\rm rad}\le 2P_{\rm opt}/c\), thermal dummy response at equal dissipated power, bearing axial thrust measured with LED OFF, and aerodynamic response measured in a sealed chamber or reduced pressure. A candidate result is accepted only if \(R_{\rm odd}\) exceeds both \(F_{\rm min}\) and the independently bounded sum of these conventional channels.

\section{Progressive validation ladder}
The experiment is not presented as a one-step demonstration of a gravitational anomaly or of the full extended Aharonov-Bohm framework. It is a staged falsification and validation protocol. The first measurement searches for a controlled apparent weight signature; subsequent tests decide whether the signature can be attributed to the optical Observer-Split transduction hypothesis rather than to ordinary forces.

\begin{table}[H]
\centering
\small
\begin{tabular}{>{\raggedright\arraybackslash}p{0.18\linewidth}>{\raggedright\arraybackslash}p{0.42\linewidth}>{\raggedright\arraybackslash}p{0.30\linewidth}}
\toprule
Stage & Required test & Interpretation if passed \\
\midrule
Candidate signature & LED ON/OFF response with CW/CCW sign reversal & Shows the modeled reversible weight signature. \\
Angular control & Localized 10--20 degree sector, sector relocated relative to the rotor, and uniform 360 degree illumination & Tests the role of \(\partial_\phi\rho_{\rm ind,Al}\). \\
Material control & Aluminum disc compared with plastic or insulating carrier of similar geometry & Tests whether conductive metallic transduction is required. \\
Optical coupling & Ordinary side-glow fiber compared with tapered, side-polished, or true evanescent-coupler sections & Tests whether the response follows evanescent coupling rather than visible scattering. \\
Artifact rejection & Calibrated thermal dummy, bearing axial thrust check, vibration monitoring, aerodynamic shielding or vacuum, magnetic/electric checks & Excludes conventional force channels quantitatively. \\
Geometric reversal & Apparatus orientation reversal with multiple independent force sensors & Separates vertical force from support-frame or bearing artifacts. \\
Blind acquisition & Pseudo-random optical switching, rotor-phase synchronous acquisition, and preregistered statistical model & Prevents operator and selection bias. \\
Scaling tests & Variation with \(\Omega\), optical power, wavelength, sector width, and orientation & Tests whether the response follows the proposed source coordinate. \\
Independent replication & Repetition by an external laboratory using the same preregistered controls & Promotes a candidate anomaly to high-confidence evidence. \\
\bottomrule
\end{tabular}
\caption{Validation ladder from a first candidate weight signature to high-confidence experimental support.}
\end{table}

If the full ladder is satisfied, including quantitative exclusion of thermal, vibrational, aerodynamic, bearing-related, electromagnetic, and optical-pressure channels, the result would first establish a reproducible optical-rotational anomalous force signature. Attribution to the extended Aharonov-Bohm Observer-Split mechanism would require the additional agreement of the calibrated \(\chi_{\rm Al}\), the response coefficient \(\kappa_{\rm opt}\), and the measured scaling with \(\Omega\), optical power, sector width, material, and orientation. Only after these conditions and independent replication are satisfied would the interpretation rise to high-confidence support for the optical Observer-Split transduction mechanism. In operational terms it could then be described as a controlled anomalous gravitational response. If one or more decisive controls fail, the corresponding conventional or apparatus-specific explanation must be preferred.

\section{From device anomaly to gravitational attribution}
The minimal rotating-disc test can establish a controlled optical-rotational anomalous vertical force only at the device level. A stronger gravitational interpretation requires a second attribution stage in which the rotating optical source acts on external test masses mechanically separated from the rotor, support frame, bearings, air flow, and optical/electrical hardware. This stage tests whether the signal is merely an apparatus force or whether it behaves as an interaction field.

The attribution chain is:
\begin{equation}
\begin{aligned}
&\hbox{disc-level ON/OFF and CW/CCW signature}\\
&\quad\Rightarrow\hbox{reproducible optical-rotational anomalous force}\\
&\quad\Rightarrow\hbox{effect on external mechanically separated test masses}\\
&\quad\Rightarrow\hbox{free-fall and distance-law verification}\\
&\quad\Rightarrow\hbox{possible anomalous gravitational interpretation}\\
&\quad\Rightarrow\hbox{quantitative compatibility with extended Aharonov-Bohm Observer-Split scaling.}
\end{aligned}
\end{equation}

The decisive external-mass tests are the following. First, a passive test mass placed above, below, or laterally displaced from the rotor should show the same optical ON/OFF and CW/CCW-odd signature without mechanical contact. Second, a freely falling or pendulum-suspended mass should show a corresponding acceleration or deflection, not merely a change in the load cell supporting the rotor. Third, the response should be independent of electric charge, magnetic susceptibility, conductivity, and material composition of the test mass, within experimental uncertainty. Fourth, the response should survive vacuum operation, electromagnetic shielding, optical baffling, and thermal isolation. Fifth, the signal should follow a reproducible distance, orientation, and source-strength law that can be fitted by the same calibrated parameters \(\chi_{\rm Al}\), \(\kappa_{\rm opt}\), and \(\Lambda_{\rm opt}\) extracted from the source-disc experiment.

The following operational criterion is therefore proposed:
\begin{equation}
\begin{aligned}
\mathcal{G}_{\rm attr}=1
\quad\Longleftrightarrow\quad
&
R_{\rm odd}^{\rm disc}>F_{\rm min},\\
&
R_{\rm odd}^{\rm ext}>F_{\rm min}^{\rm ext},\\
&
a_{\rm test}^{\rm ON}-a_{\rm test}^{\rm OFF}\neq0,\\
&
\hbox{test-mass material independence},\\
&
\hbox{survival under shielding and vacuum},\\
&
\hbox{agreement with calibrated Observer-Split scaling}.
\end{aligned}
\end{equation}
The binary marker \(\mathcal{G}_{\rm attr}\) is not a new field variable; it is a pre-registered attribution flag that is set to one only when all listed tests are passed.
Here \(R_{\rm odd}^{\rm ext}\) is the odd ON/OFF response of an external test mass and \(a_{\rm test}\) is its measured acceleration or pendulum-equivalent acceleration. This criterion is intentionally stronger than the minimal disc test. Passing the disc test would establish a candidate anomalous force. Passing the external-mass and free-fall tests would promote the result to convergent evidence for a controlled anomalous gravitational interaction. Passing the same tests with the calibrated Observer-Split scaling would then support the extended Aharonov-Bohm interpretation rather than only the existence of an unknown vertical force.

\section{Discussion and limitations}
The model is deliberately ambitious but falsifiable. Its central novelty is replacing a rotating magnetic source proxy by an optical gate that modulates an aluminum-induced effective density while keeping the observer-split source law unchanged. This avoids the unjustified direct identification \(T_{\rm ev}\equiv\rho\): \(T_{\rm ev}\) is a gate, whereas \(\rho_{\rm ind,Al}\) is the phenomenological density entering \(I_{\rm opt}=\Omega\partial_\phi\rho_{\rm ind,Al}\).

The optical route is attractive because it is light, spatially maskable, rapidly switchable, and free of direct magnetic attraction artifacts. The principal risk is that an ordinary optical side-glow fiber may not produce a sufficiently strong evanescent or surface-coupling channel in aluminum. A future improved version should use tapered, side-polished, or true evanescent coupler fiber sections, and should independently calibrate the optical-to-aluminum coefficient \(\chi_{\rm Al}\). In the absence of such calibration, a null result constrains the product \(|\kappa_{\rm opt}\chi_{\rm Al}|\) for the tested geometry rather than falsifying the general Observer-Split identity.

The two constitutive assumptions are deliberately exposed. First, \(J^\mu_{\rm ind,Al}\) is the effective conserved aluminum response current inferred from independently measured optical-to-metal modulation. Second, \(K_z\), or equivalently the reduced coefficient \(\kappa_{\rm opt}\), is the apparatus response that maps the extended scalar source into a vertical residual force. The experiment is successful as physics even in the null case if it turns these quantities into numerical upper bounds. It becomes positive evidence for the optical Observer-Split branch only if the same calibrated coefficients predict the sign, scaling, material dependence, and orientation dependence across independent runs.

\section{Conclusion}
We have formulated an optically gated aluminum-induced Observer-Split source coordinate for a rotating aluminum disc. The construction integrates evanescent/tunneling optical control with the extended Aharonov-Bohm scalar closure while applying the Observer-Split term to an induced effective density, not directly to an optical transmission coefficient. The predicted signature is ON/OFF controlled, odd under CW/CCW reversal, suppressed for a uniform luminous ring, and material-dependent through \(\chi_{\rm mat}\).

The purpose of the experiment is not only to search for an immediate weight anomaly, but to provide a falsifiable path from a first optical-rotational signature to a calibrated test of the extended Aharonov-Bohm Observer-Split framework. Before calibration, the correct quantitative prediction is not a fixed force in newtons but a declared sensitivity target \(F_{\rm min}\) and a corresponding bound or estimate for \(\Lambda_{\rm opt}=\kappa_{\rm opt}\chi_{\rm Al}\). After independent density calibration, the same data separate \(\chi_{\rm Al}\) from \(\kappa_{\rm opt}\). A first positive result would be a candidate anomalous weight signature. A result that also survives the full validation ladder, independent replication, and quantitative agreement with the calibrated \(\chi_{\rm Al}\), \(\kappa_{\rm opt}\), and scaling laws would provide strong experimental support for the optical Observer-Split transduction mechanism. If the extended external-mass campaign further demonstrates material-independent action on mechanically separated test bodies and a corresponding free-fall or pendulum acceleration under shielding and vacuum controls, the interpretation would advance from device-level anomaly to strong evidence for a controlled anomalous gravitational interaction. If the predicted ON/OFF, CW/CCW, material, angular, external-mass, and artifact-rejection conditions fail, the optical source branch or its gravitational attribution is falsified for the tested geometry.

\section*{Use of generative AI language tools}
Text-to-text generative AI language tools were used for language editing, editorial assistance, and manuscript formatting. The authors reviewed the resulting text and take full responsibility for the scientific content.

\section*{Appendix A: Synthetic machine-learning consistency audit}
A synthetic neural-network audit was performed to test whether the proposed signature is separable from declared false-positive classes. The positive generator encoded \(F_z^{\rm res}\propto\Delta G_{\rm opt}\) with sign reversal and optical gating. Negative generators included null response, thermal light-only drift, vibration proportional to rpm squared, wrong-sign no-reversal response, uniform optical source, and dummy-active artifacts.

The neural-network audit does not constitute physical validation of the effect. It is used only as an internal consistency and classification check showing that the modeled Observer-Split signature is mathematically separable from the synthetic classes selected by the modeler. The neural network achieved 94.2 percent accuracy, 93.8 percent recall, and 94.7 percent specificity on held-out synthetic cases. The Roshchin-Godin benchmark reconstruction returned 35.0 percent at the encoded historical point. These results show internal mathematical separability; they are not experimental proof.

\begin{figure}[H]
\centering
\includegraphics[width=0.78\linewidth]{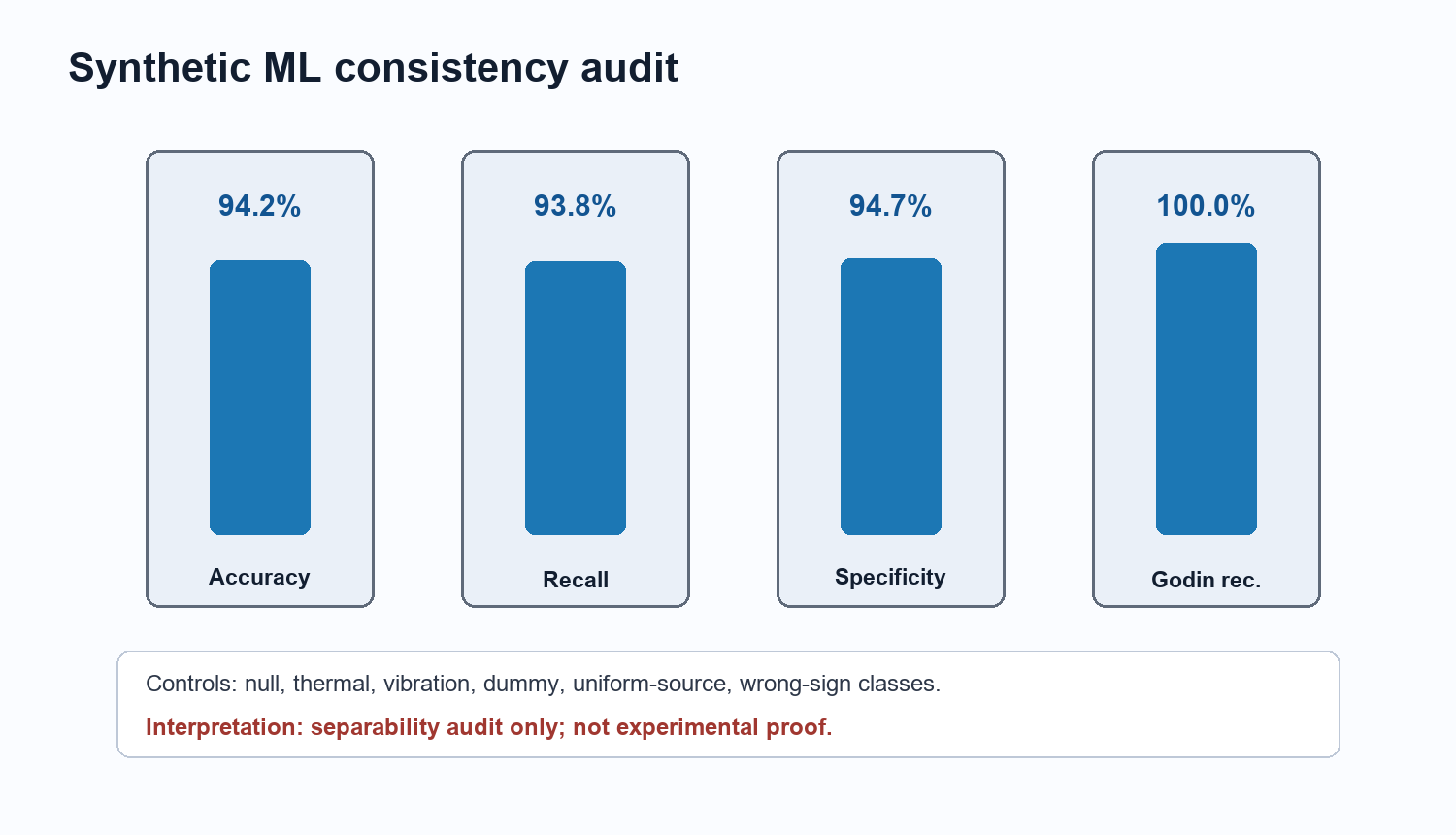}
\caption{Synthetic consistency audit. The percentages summarize classification on synthetic modeled classes only and are not experimental validation.}
\end{figure}

\section*{Code Availability}
The Python code used to generate the analytical figures, the optical Observer-Split source model, and the synthetic machine-learning consistency audit is archived on Zenodo \cite{iadicicco2026code}:
\begin{center}
\url{https://doi.org/10.5281/zenodo.21781822}
\end{center}


\clearpage
\begin{thebibliography}{99}
\bibitem{iadicicco2026split} A. Iadicicco, G. Modanese, and L. Verolino, ``Effective Observer-Split Source Terms in Rotating Frames and Gravitomagnetic Backgrounds in Extended Aharonov-Bohm Electrodynamics,'' arXiv:2604.24787 [physics.gen-ph] (2026).
\bibitem{iadicicco2026code} A. Iadicicco, ``Optical-Fiber Observer-Split Source Model and Synthetic Consistency Audit,'' Zenodo, version 1.0.0, 2026. doi:10.5281/zenodo.21781822.
\bibitem{roshchin2000} V. V. Roshchin and S. M. Godin, ``An Experimental Investigation of the Physical Effects in a Dynamic Magnetic System,'' \emph{Technical Physics Letters}, vol. 26, no. 12, pp. 1105--1107, 2000. doi:10.1134/1.1337268.
\bibitem{roshchin2004} V. V. Roschin and S. M. Godin, ``Orbiting Multi-Rotor Homopolar System,'' US Patent US6822361B1, 2004.
\bibitem{aharonov1959} Y. Aharonov and D. Bohm, ``Significance of Electromagnetic Potentials in the Quantum Theory,'' \emph{Physical Review}, vol. 115, no. 3, pp. 485--491, 1959. doi:10.1103/PhysRev.115.485.
\bibitem{aharonov1961} Y. Aharonov and D. Bohm, ``Further Considerations on Electromagnetic Potentials in the Quantum Theory,'' \emph{Physical Review}, vol. 123, pp. 1511--1524, 1961. doi:10.1103/PhysRev.123.1511.
\bibitem{chambers1960} R. G. Chambers, ``Shift of an Electron Interference Pattern by Enclosed Magnetic Flux,'' \emph{Physical Review Letters}, vol. 5, pp. 3--5, 1960. doi:10.1103/PhysRevLett.5.3.
\bibitem{simmons1963} J. G. Simmons, ``Generalized Formula for the Electric Tunnel Effect between Similar Electrodes Separated by a Thin Insulating Film,'' \emph{Journal of Applied Physics}, vol. 34, pp. 1793--1803, 1963. doi:10.1063/1.1702682.
\bibitem{hartman1962} T. E. Hartman, ``Tunneling of a Wave Packet,'' \emph{Journal of Applied Physics}, vol. 33, pp. 3427--3433, 1962. doi:10.1063/1.1702424.
\bibitem{steinberg1993} A. M. Steinberg, P. G. Kwiat, and R. Y. Chiao, ``Measurement of the Single-Photon Tunneling Time,'' \emph{Physical Review Letters}, vol. 71, pp. 708--711, 1993. doi:10.1103/PhysRevLett.71.708.
\bibitem{balcou1997} P. Balcou and L. Dutriaux, ``Dual Optical Tunneling Times in Frustrated Total Internal Reflection,'' \emph{Physical Review Letters}, vol. 78, pp. 851--854, 1997. doi:10.1103/PhysRevLett.78.851.
\bibitem{longhi2005} S. Longhi, ``Resonant Tunneling in Frustrated Total Internal Reflection,'' \emph{Optics Letters}, vol. 30, no. 20, pp. 2781--2783, 2005. doi:10.1364/OL.30.002781.
\bibitem{philbin2008} T. G. Philbin, C. Kuklewicz, S. Robertson, S. Hill, F. Konig, and U. Leonhardt, ``Fiber-Optical Analog of the Event Horizon,'' \emph{Science}, vol. 319, no. 5868, pp. 1367--1370, 2008. doi:10.1126/science.1153625.
\bibitem{modanese2017} G. Modanese, ``Generalized Maxwell Equations and Charge Conservation Censorship,'' \emph{Modern Physics Letters B}, vol. 31, 1750052, 2017. doi:10.1142/S021798491750052X.
\bibitem{minotti2021} F. Minotti and G. Modanese, ``Quantum Uncertainty and Energy Flux in Extended Electrodynamics,'' \emph{Quantum Reports}, vol. 3, pp. 703--723, 2021. doi:10.3390/quantum3040044.
\bibitem{ruggiero2005} M. L. Ruggiero, ``The Sagnac Effect in Curved Space-Times from an Analogy with the Aharonov-Bohm Effect,'' \emph{General Relativity and Gravitation}, vol. 37, pp. 1845--1855, 2005. doi:10.1007/s10714-005-0190-0.
\end{thebibliography}
\end{document}